\documentclass[aps,twocolumn,showpacs,floatfix]{revtex4}
\usepackage{amsmath}
\usepackage{amsfonts}
\usepackage{amssymb}
\usepackage{graphicx}
\usepackage{bbm}

\begin{document}
\title{Amplification of NOON States}
\author{G. S. Agarwal,$^1$ S. Chaturvedi,$^2$ and Amit Rai$^1$}
\affiliation{1 Department of Physics, Oklahoma State University,
Stillwater, OK 74078, USA}

\affiliation{2 School of Physics, University of Hyderabad, Hyderabad
500046, India}

\begin{abstract}
We examine the behavior of a Non Gaussian state like NOON state under phase insensitive amplification. We derive analytical result
for the density matrix of the NOON state for arbitrary gain of the amplifier. We consider cases of both symmetric and antisymmetric
amplification of the two modes of the NOON state. We quantitatively evaluate the loss of entanglement by the amplifier in terms
of the logarithmic negativity parameter. We find that NOON states are more robust than their Gaussian counterparts.
\end{abstract}

\pacs{42.50.Ex, 42.50.Nn, 03.65.Ud }

\maketitle
\section{Introduction}

\noindent Among the continuous variable (CV) entangled states \cite{1}, non Gaussian states are generally believed
to possess much more robust entanglement vis a vis the Gaussian states-- states characterized by a Gaussian
 quasi probability distributions and hence by their first and second moments. Though  mathematically not as
 well understood  as the Gaussian states \cite{2} in so far as their entanglement properties are concerned,
 non Gaussian states, by virtue of the robustness of their entanglement have in recent years emerged as strong
 contenders for potential applications in quantum information technology.
The fact that the non Gaussian states are defined by what they are
not makes a general discussion of their entanglement properties
impossible and one is forced to restrict oneself to sub families of
non Gaussian states such as states obtained by adding or subtracting
a fixed number of photons to Gaussian states \cite{3} and forming
suitable superpositions thereof. One such widely discussed family of
non Gaussian states parameterized by an integer $N$ is the family of
NOON states \cite{4}

\begin{equation}
 |NOON>=\frac{1}{\sqrt{2}}~[|N,0>+|0,N>].
\end{equation}

\noindent These states, similar in structure to the EPR states, have attracted
much attention in recent years and can be viewed as a two mode state
consisting of a superposition of states containing $N$ photons in one
mode and none in the other and vice versa. Schemes for reliable
production of such states have been proposed \cite{5, Vitelli}  and their
usefulness as a practical tool in making super-precision measurements
in optical interferometry, atomic spectroscopy  and in sensing
extremely small magnetic fields  than hitherto possible have been
highlighted \cite{6}. This circumstance makes it imperative to
investigate their behavior under attenuation and amplification.
While studies on the decoherence effects on NOON states under
specific models for system-bath interactions already exist in the
literature \cite{7}, in the present work we focus on the question of
amplification of NOON states and the consequent degradation of
entanglement therein and to compare and contrast it with the
behavior of entanglement in Gaussian states under amplification
investigated in \cite{8, 9}. It is important to understand the
amplification of NOON states as quantum communication protocols
would use both amplifiers and attenuators \cite{Yuen}. In the
present work we consider only phase insensitive amplification  and
model this in the standard way as a bath consisting of $N$
two level atoms of which $N_1$ are in the excited state and $N_2$
in the ground state with $N_1>N_2$. Under the assumptions that
atomic transitions have a large width and that the bath is
maintained in a steady state, the time evolution of the density
operator $\rho$ for a single mode of radiation field on resonance
with the atomic transition is described, in the interaction
picture,  by the master equation

\begin{eqnarray}
\frac{\partial\rho}{\partial t}&=&-\kappa
N_1(aa^\dagger\rho-2a^\dagger\rho a+\rho  aa^\dagger)\nonumber\\
&&-\kappa N_2(a^\dagger a \rho-2a\rho a^\dagger+\rho a^\dagger a),
\label{c}
\end{eqnarray}

\noindent where $a$ and $a^\dagger$ are the annihilation and creation
operators of the field mode.

\noindent A brief outline of this work is as follows. In Section II,
with the NOON state as the input to the amplifier, we obtain
expressions for the output density operator for the case when both
the modes are symmetrically amplified and for the case when only one
mode is subjected to amplification and the other is not amplified at
all. We investigate how the entanglement in the output state varies
with the amplifier gain using logarithmic negativity as the
quantifier for entanglement \cite{10}. Section III contains our
concluding remarks and further outlook. Our study complements the
work done by Vitelli et al \cite{Vitelli} on amplification of NOON
state by a phase sensitive amplifier.

\section{Evolution of the NOON state under phase insensitive amplification}

\noindent In our earlier work \cite{8},  we found that for a two mode squeezed vacuum as the
input, there are limits on the gain beyond which the output of the amplifier has no
entanglement between the two modes and the limiting values of the gain  in the two
cases considered,  symmetric and asymmetric amplification, were   found to be

\begin{eqnarray}
 G^2&=&\left(\frac{2+2\eta}{1+2\eta+e^{-2r}}\right) {\rm Symmetric~Amplification}~~\\
G^2&=& 1+ \frac{1}{\eta}~~~~~~~~{\rm Asymmetric ~Amplification}.
\end{eqnarray}

\noindent where the gain $G = \exp((N_1-N_2)\kappa t)$ and
$\eta=N_2/(N_2-N_1)$. In particular, when $\eta\rightarrow 0$, the
limiting gain in the symmetric case remains finite, that for the
asymmetric case the limiting value moves off to infinity \cite{Pooser}.

\noindent In this section we discuss how an input NOON state evolves under the action of a
phase insensitive amplifier as modeled by the master equation $(\ref{c})$ confining
ourselves for simplicity to the $\eta\rightarrow 0$ limit. The solution of master equation (2) can be
written in terms of the Fock state matrix elements. However such a solution is rather
involved \cite{Chaturvedi}. It is instructive to work in terms of phase space
distributions. A distribution which is especially useful is the Q function introduced by Kano; Sudarshan and Mehta \cite{Kano, Sudarshan}. This
function is defined by

\begin{equation}
  Q(\alpha) \equiv \frac{1}{\pi}<\alpha|\rho|\alpha>
\end{equation}

\noindent The master equation (2) for $\eta=0$ then leads to

\begin{equation}
\frac{\partial Q}{\partial t} = -G \frac{\partial (\alpha
Q)}{\partial \alpha}-G \frac{\partial (\alpha^{*} Q)}{\partial
\alpha^{*}}
\end{equation}

\noindent Note that this differential equation for $Q$ function involves only
the first order derivatives with respect to phase space variables
and hence its solution is simple \cite{gsa}

\begin{equation}
  Q_{in}(\alpha) \equiv \frac{1}{\pi}<\alpha|\rho_{in}|\alpha> \rightarrow Q_{out}(\alpha)=\frac{1}{G^2}Q_{in}(\alpha/G)
\label{d}
\end{equation}

\noindent Let us see what the result $(\ref{d})$ means. Let us consider the input state to be vacuum then the output would be

\begin{eqnarray}
 Q_{in}(\alpha)  & = & \frac{1}{\pi} <\alpha|0> <0|\alpha> \nonumber\\
& = &   \frac{1}{\pi} e^{-|\alpha|^2} \rightarrow Q_{out}(\alpha) \equiv \frac{1}{\pi G^2}
e^{-|\alpha|^2/ G^2}
\label{7}
\end{eqnarray}

\noindent Such an output $Q$ function is equivalent to a thermal
density matrix with mean number of photons equal to $(G^2-1)$. Thus the vacuum state on amplification becomes a thermal
state with a mean number of photons that grows with the gain of the amplifier. We note
that a result like (6) extends to multi-mode case.

\noindent We now examine two cases, the symmetric case in which both
the modes $a$ and $b$ in the input NOON state are symmetrically
amplified and the asymmetric case in which only one mode, say $a$,
is amplified. Before discussing the amplification of the NOON state
we examine quantitatively the entanglement in the state (1). We
compute the log negativity parameter which is defined as

\begin{eqnarray*}
E_{N}=\log_{2}(2N+1)~,
\end{eqnarray*}

\noindent where $N$ is the absolute value of the sum of all the
negative eigenvalues of the partial transpose of the density matrix $\rho$. It is clear
that the partial transpose of the density matrix associated with the
state (1) is

\begin{eqnarray}
 \rho^{pt}&=&\frac{1}{2}[|N,0><N,0|+|0,N><0,N|\nonumber\\
&&+|N,N><0,0|+|0,0><N,N|]~,
\end{eqnarray}

\noindent which can be written in the diagonal form as

\begin{eqnarray}
 \rho^{pt} &=& \frac{1}{2}(|N,0><N,0|+|0,N><0,N|)\nonumber\\
&&+\frac{1}{4}(|N,N>+
 |0,0>)(<N,N|+<0,0|)-\nonumber\\&&\frac{1}{4}(|N,N>-|0,0>)(<N,N|-<0,0|)
\end{eqnarray}

\noindent The partial transpose has a negative eigenvalue $-1/2$ and
hence the logarithmic negativity parameter $E_{N}=1$.

\vskip5mm

\noindent {\it Symmetric case}: \noindent The $Q$- function
corresponding to  the density operator for the input NOON state

\begin{eqnarray}
 \rho_{in}&=&\frac{1}{2}[|N,0><N,0|+|N,0><0,N|\nonumber\\&&
 +|0,N><N,0|+|0,N><0,N|]\nonumber\\
&=&\frac{1}{2N!}[a^{\dagger N}\rho_0a^N+a^{\dagger N}\rho_0b^N+b^{\dagger N} \rho_0a^N
+b^{\dagger N}\rho_0b^N],\nonumber\\~~
\rho_0 &=& |0,0><0,0|~,
\end{eqnarray}

\noindent is found to be

\begin{eqnarray}
 Q_{in}(\alpha,\beta)&\equiv&\frac{1}{\pi^2}<\alpha,\beta|\rho_{in}|\alpha,\beta>\nonumber\\
&=& \frac{1}{2N!\pi^2}|\alpha^N+\beta^N|^2\exp[-(|\alpha|^2+|\beta|^2)]\nonumber\\
\end{eqnarray}

\noindent Following the prescription in $(\ref{d})$, under symmetric phase insensitive amplification, the Q-function evolves as follows:

\begin{eqnarray}
 Q_{in}(\alpha,\beta)\rightarrow Q_{out}(\alpha,\beta)&=&\frac{1}{G^4}Q_{in}(\alpha/G,\beta/G)\nonumber\\
&=&\frac{1}{2N!\pi^2 G^{2N+4}}|\alpha^N+\beta^N|^2\nonumber\\&&\times\exp[-(|\alpha|^2+|\beta|^2)/G^2]\nonumber\\
\label{12}
\end{eqnarray}

\noindent The NOON state is highly nonclassical. A quantitative measure for nonclassicality is obtained by examining zeroes of the $Q$ function \cite{wolf}. We note that the zeroes of the function $Q_{out}$ are
identical to the zeroes of the function $Q_{in}$ and hence we have
the remarkable result that the nonclassical character of the input
NOON state is preserved.

\noindent We can now find the density matrix after amplification by
using the results (\ref{7})  and (\ref{12}) :

\begin{eqnarray}
 \rho_{in}\rightarrow \rho_{out}&=&\frac{1}{2N!G^{2N}}[a^{\dagger N}\rho_{G}a^N+a^{\dagger N}\rho_{{\rm G}}b^N\nonumber\\
&&+
 b^{\dagger N}\rho_{G}a^N+b^{\dagger N}\rho_{G}b^N],\nonumber\\
\rho_{G}&=&\frac{1}{G^4}e^{-\beta(a^\dagger a+b^\dagger b)},\nonumber\\
&& \nonumber\\~\beta &=& {\rm ln}\left(\frac{G^2}{G^2-1}\right).
\label{density}
\end{eqnarray}

\noindent We note that the structure of (\ref{density}) is such that
it can not be written in a separable form. This is seen more clearly
if we write (\ref{density}) as

\begin{eqnarray}
 \rho_{out}= \frac{1}{2N!G^{2N}}\{a^{\dagger N}+ b^{\dagger
 N}\}\rho_{G}\{a^{ N}+ b^{N}
\}
 \end{eqnarray}

\noindent We further note that the output state has the structure of
a two mode photon added thermal state in which either mode has added
photons. The single mode version of photon added thermal state was
introduced by Agarwal and Tara \cite{tara}. These states have been
experimentally studied recently \cite{Bellini}.

\noindent Writing $\rho_{G}$ in the number state basis as

\begin{equation}
\rho_{G} = \frac{1}{G^4}\sum_{n,m=0}^{\infty}~\left(\frac{G^2-1}{G^2}\right)^{n+m}|n,m><n,m|
\end{equation}

\noindent we have

\begin{eqnarray}
\rho_{out}&=&\frac{1}{2N!G^{2N+4}}\sum_{n,m=0}^{\infty}\left(\frac{G^2-1}{G^2}\right)^{n+m}
\nonumber\\
&&\times[a^{\dagger N}|n,m><n,m|a^N+a^{\dagger
N}|n,m><n,m|b^N\nonumber\\&&+b^{\dagger N}|n,m><n,m|a^N+b^{\dagger
N}
|n,m><n,m|b^N]\nonumber\\
&=&\frac{1}{2N!G^{2N+4}}\sum_{n,m=0}^{\infty}\left(\frac{G^2-1}{G^2}\right)^{n+m}\nonumber\\
&&\times[\frac{(n+N)!}{n!}
|n+N,m><n+N,m|\nonumber\\
&&+\frac{(m+N)!}{m!}
|n,m+N><n,m+N|\nonumber\\
&&+\sqrt{\frac{(n+N)!}{n!}\frac{(m+N)!}{m!}}(|n+N,m><n,m+N|\nonumber\\
&&+|n,m+N><n+N,m|)]
\end{eqnarray}

\noindent which immediately gives us the expression for the operator
$\rho_{out}^{PT}$ obtained by partially transposing $\rho_{out}$ (
with respect to the b mode):

\begin{eqnarray}
\rho_{out}^{PT}&=&\frac{1}{2N!G^{2N+4}}\sum_{n,m=0}^{\infty}\left(\frac{G^2-1}{G^2}\right)^{n+m}\nonumber\\
&&\times[\frac{(n+N)!}{n!}
|n+N,m><n+N,m|\nonumber\\
&&+\frac{(m+N)!}{m!}
|n,m+N><n,m+N|\nonumber\\
&&+\sqrt{\frac{(n+N)!}{n!}\frac{(m+N)!}{m!}}(|n+N,m+N><n,m|\nonumber\\
&&+|n,m><n+N,m+N|)]\nonumber\\
&=&\frac{1}{2N!G^{2N}}[a^{\dagger N}\rho_{G}a^N+b^{\dagger N}\rho_{G}b^N \nonumber\\
&&+a^{\dagger N}b^{\dagger N} \rho_{G}+\rho_{G}a^N b^N]
\end{eqnarray}

\begin{figure}[htp]
 \scalebox{0.48}{\includegraphics{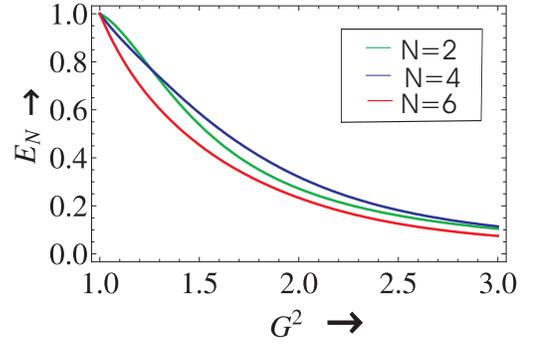}}
 \caption{\label{Fig1(a)} Behavior of the logarithmic negatively as a function of $G^2$ for the symmetric case.}
\end{figure}

\noindent The object of interest now is to calculate the logarithmic
negativity $E_\mathcal{N}$, the sum of the logarithmic negativity
eigen-values of $\rho_{out}^{PT}$ and to see how it varies as a
function of $G^2$. We carry out this task numerically and the
results are displayed in Fig. 1 where we plot $E_\mathcal{N}$ as a
function of $G^2$ for $N=2,4$ and 6.

\vspace{0.1 cm}

\noindent {\it Asymmetric case}: Proceeding as above, one finds that

\begin{eqnarray}
 Q_{in}(\alpha,\beta)\rightarrow Q_{out}(\alpha,\beta)&=&\frac{1}{G^2}Q_{in}(\alpha/G,\beta)\nonumber\\
&=&\frac{1}{2N!\pi^2 G^{N+2}}|(\alpha/G)^N+\beta^N|^2\nonumber\\&&
\times \exp[-(|\alpha|^2/G^2+|\beta|^2)]
\end{eqnarray}

\noindent and hence

\begin{eqnarray}
 \rho_{in}\rightarrow \rho_{out}&=&\frac{1}{2N!G^{2N}}[a^{\dagger N}\tilde{\rho}a^N
 +G^N a^{\dagger N}\tilde{\rho}b^N\nonumber\\&&+G^N
 b^{\dagger N}\tilde{\rho}a^N+G^{2N}b^{\dagger N}\tilde{\rho}b^N]\nonumber\\
\tilde{\rho} &=& \frac{1}{G^2}e^{-\beta(a^\dagger a)}|0><0|;~\beta={\rm ln}\left(\frac{G^2}{G^2-1}\right)\nonumber\\
\end{eqnarray}

\begin{figure}[htp]
 \scalebox{0.48}{\includegraphics{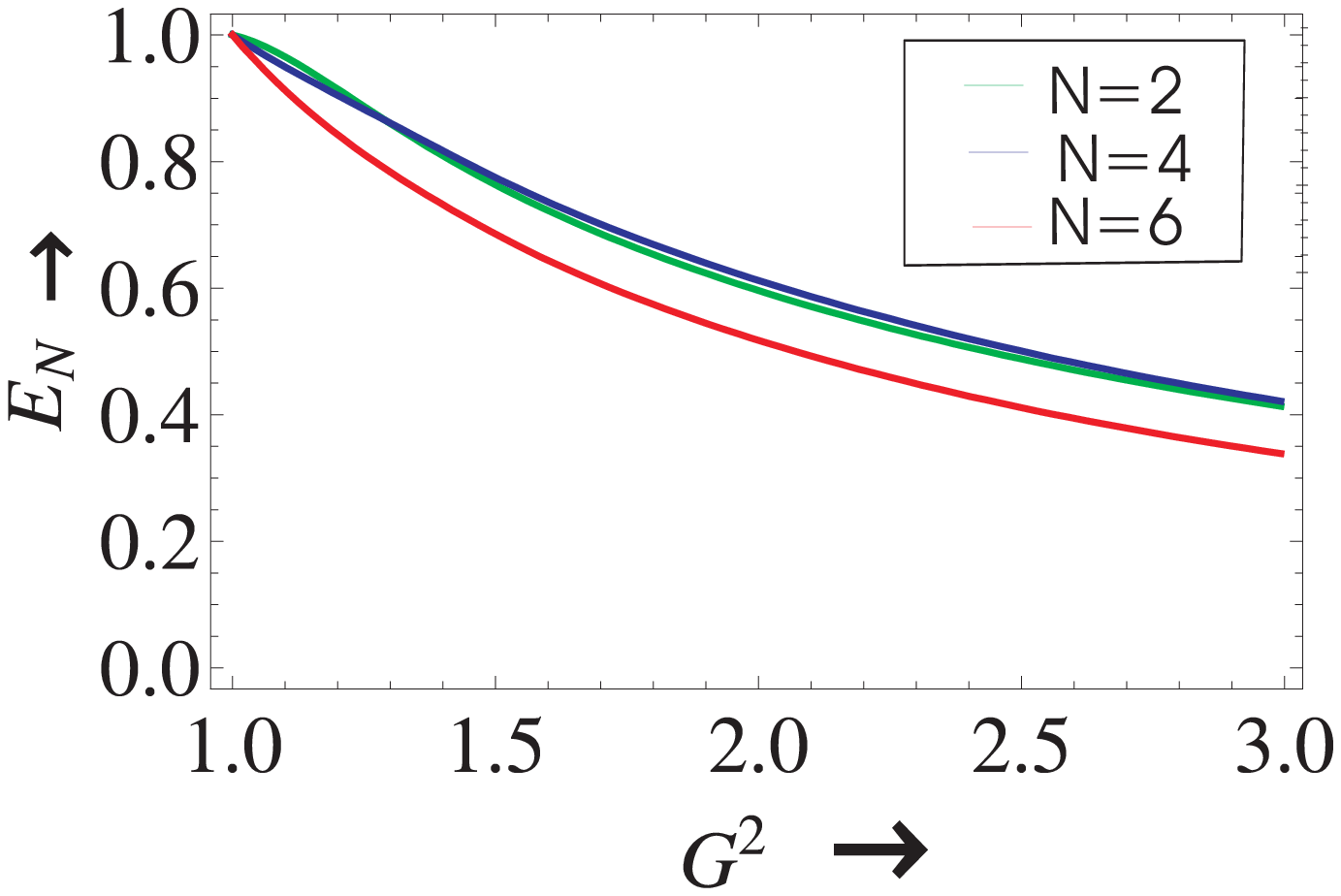}}
 \caption{\label{Fig1(a)} Behavior of the logarithmic negatively as a function of $G^2$ for the asymmetric case.}
\end{figure}

\noindent Writing $\tilde{\rho}$ as

\begin{equation}
\tilde{\rho}= \frac{1}{G^2}\sum_{n=0}^{\infty}~\left(\frac{G^2-1}{G^2}\right)^{n}|n,0><n,0|
\end{equation}

\noindent we can write $\rho_{out}$ in terms of number states as

\begin{eqnarray}
\rho_{out}&=&\frac{1}{2N!G^{2N+2}}\sum_{n=0}^{\infty}\left(\frac{G^2-1}{G^2}\right)^{n}
\nonumber\\
&&\times[a^{\dagger N}|n,0><n,0|a^N+G^N a^{\dagger
N}|n,0><n,0|b^N\nonumber\\&&+G^N b^{\dagger
N}|n,0><n,0|a^N\nonumber\\
&&+G^{2N}b^{\dagger N}
|n,0><n,0|b^N]\nonumber\\
&=&\frac{1}{2N!G^{2N+2}}\sum_{n=0}^{\infty}\left(\frac{G^2-1}{G^2}\right)^{n}\nonumber\\
&&\times[\frac{(n+N)!}{n!} |n+N,0><n+N,0|\nonumber\\
&&+G^{2N} N!
|n,N><n,N|\nonumber\\
&&+G^N\sqrt{\frac{(n+N)!}{n!}N!}(|n+N,0><n,N|\nonumber\\
&&+|n,N><n+N,0|)]
\end{eqnarray}

\noindent and hence

\begin{eqnarray}
\rho_{out}^{PT}&=&\frac{1}{2N!G^{2N+2}}\sum_{n=0}^{\infty}\left(\frac{G^2-1}{G^2}\right)^{n}\nonumber\\
&&\times[\frac{(n+N)!}{n!} |n+N,0><n+N,0|\nonumber\\
&&+G^{2N} N! |n,N><n,N|\nonumber\\
&& + G^N \sqrt{\frac{(n+N)!}{n!}N!} (|n+N,N><n,0|\nonumber\\
&&+|n,0><n+N,N|)]\label{b}\nonumber\\
&=&\frac{1}{2N!G^{2N}}[a^{\dagger N}\tilde{\rho}a^N+G^{2N}b^{\dagger N}\tilde{\rho}b^N\nonumber\\
&&+G^N a^{\dagger N}b^{\dagger N} \tilde{\rho}+G^N \tilde{\rho}a^N
b^N]
\end{eqnarray}

\vspace{1.5 cm}

\noindent The logarithmic negativity of $\rho_{out}^{PT}$ is
computed numerically and the results are shown in Fig. 2 for
$N=2,4,6$. Clearly in the asymmetric case the loss of entanglement
is much slower. Finally we compare the amplification of the NOON
state with that of photon added two mode squeezed vacuum state i.e.
the state (for brevity the normalization factors are ignored)

\begin{equation}
|\Phi\rangle \propto a^{\dagger} b^{\dagger} \exp\{\zeta a^{\dagger}
b^{\dagger} -\zeta^{*} a b \}|0 0\rangle
\end{equation}

\noindent This is a non Gaussian state. The input and output $Q$
functions are found to be

\begin{equation}
Q_{in}   \propto  |\alpha |^2 |\beta |^2 Q_{sq}
\end{equation}

\begin{equation}
Q_{out} \propto   \frac{|\alpha |^2 |\beta |^2}{G^4} Q_{sq,G}
\end{equation}

\noindent where $Q_{sq}$ is the $Q$ function for the two mode
squeezed vacuum and $Q_{sq,G}$ is the $Q$ function obtained by
amplification of the squeezed vacuum. Thus the density operator
after amplification of the non Gaussian state can be written as

\begin{equation}
\rho_{out} \propto a^{\dagger} b^{\dagger} (\rho_{sq out}) a b ,
\end{equation}

\noindent where $\rho_{sq out}$ is the density operator for the squeezed vacuum after amplification. Now $\rho_{sq out}$ becomes
separable for $G$ greater than that given by (3) and hence $\rho_{out}$ becomes separable if $G^2>(2+2\eta)/(1+2 \eta+e^{-2r})$.
Thus the non Gaussian states obtained from Gaussian states by the addition of photons would behave under symmetric amplification in
a manner similar to Gaussian states. We have therefore found that the NOON states behave quite differently under amplification.

\section{Conclusions}

\noindent In conclusion we have found that the NOON states are more
robust under amplification than their Gaussian counter parts such as
two mode squeezed vacuum produced by a down converter. We have
presented results for the logarithmic negativity as a function of
the gain of the amplifier. We have presented numerical results for
cases of states which already have been realized experimentally.
 We also found that the NOON state does better than say photon added two mode squeezed vacuum state.

\end{document}